\documentclass[10pt,twocolumn]{article}

\usepackage[letterpaper,margin=0.72in,columnsep=0.25in]{geometry}
\usepackage[T1]{fontenc}
\usepackage[utf8]{inputenc}
\usepackage{lmodern}
\usepackage{microtype}

\usepackage{amsmath,amssymb,amsfonts}
\usepackage{physics}
\usepackage{graphicx}
\usepackage{xcolor}
\usepackage{booktabs}
\usepackage{array}
\usepackage{caption}
\usepackage{dblfloatfix}

\usepackage[numbers,sort&compress]{natbib}
\usepackage[colorlinks=true,allcolors=blue]{hyperref}
\usepackage{cleveref}

\usepackage{titling}
\usepackage{authblk}

\crefname{figure}{Fig.}{Figs.}
\Crefname{figure}{Fig.}{Figs.}
\crefname{table}{Table}{Tables}
\Crefname{table}{Table}{Tables}
\crefname{equation}{Eq.}{Eqs.}
\Crefname{equation}{Eq.}{Eqs.}
\crefname{subsection}{section}{sections}
\Crefname{subsection}{Section}{Sections}

\title{Dynamic Entanglement Distribution for Multi-User and Multi-Protocol Quantum Networking}

\author[1,2]{Rui Wang\thanks{Corresponding author: rui.wang@bristol.ac.uk}}
\author[2,3]{Marcus J. Clark}
\author[1,2]{Obada Alia}
\author[1,2]{Sima Bahrani}
\author[2,3,4]{Djeylan Aktas}
\author[5]{Matej Perani\'{c}}
\author[5]{Mario Stipčević}
\author[5]{Martin Lončarić}
\author[2,3]{John Rarity}
\author[2,3]{Siddarth K. Joshi}
\author[1,2]{Dimitra Simeonidou}

\affil[1]{Smart Internet Lab, University of Bristol, Bristol, UK}
\affil[2]{School of Electrical, Electronic, and Mechanical Engineering, University of Bristol, Bristol, UK}
\affil[3]{Quantum Engineering Technology Labs, University of Bristol, Bristol, UK}
\affil[4]{Institute of Physics, Slovak Academy of Sciences, Dubravska cesta 9, Bratislava 845 11, Slovakia}
\affil[5]{Laboratory for Photonics and Quantum Optics, Ruder Bošković Institute, Zagreb, Croatia}

\date{}

\begin{document}

\maketitle

\begin{abstract}
Dynamic entanglement distribution is a key requirement for scalable, multi-user and multi-protocol quantum networks. We demonstrate a metropolitan-scale entanglement-based quantum communication network enabled by a quantum reconfigurable optical add-drop multiplexer (q-ROADM), which dynamically distributes polarisation-entangled photon pairs from a broadband source to six users over deployed campus and metropolitan fibre. The network supports programmable full-mesh, partial-mesh and sliced sub-network configurations, enabling flexible allocation of entanglement resources according to link condition and service requirement. We demonstrate stable six-user full-mesh operation over more than 150~hours, compare full-mesh and time-shared partial-mesh strategies under different source and detector conditions, and realise quantum network slicing with optional/additional interconnection links. We also show that the same infrastructure can support different quantum protocols by showcasing Secure Inaugural Authentication-Transfer (SIAT) combined with Network flooding over multiple paths to improve the security of onboarding a new user. These results demonstrate a q-ROADM-enabled entanglement distribution architecture as a novel route towards reconfigurable, service-oriented quantum networking over optical fibre infrastructure.
\end{abstract}

\section{Introduction}

The quantum Internet is expected to evolve from current point-to-point quantum links towards large-scale, multi-user quantum networking infrastructures~\cite{wehner2018quantum,Kimble2008QuantumInternet,Azuma2023QuantumRepeaters}. In this context, quantum networks that interconnect multiple quantum nodes are essential for enabling applications such as quantum key distribution (QKD)~\cite{ribordy2000long, yin2020entanglement,Cacciapuoti2020QuantumInternet}, quantum teleportation ~\cite{bouwmeester1997experimental,pirandola2015advances,Thomas2024TeleportationCoexistence,valivarthi2020teleportation,valivarthi2016quantum}, distributed quantum computing \cite{cirac1999distributed,main2025distributed, cacciapuoti2019quantum} and, ultimately, service-oriented quantum networking \cite{ahmed2025osi}. 
Similar to the classical Internet in providing communication services, one of the long-term goals of the quantum Internet is to offer widespread and flexible access to quantum resources (i.e., entanglement between network edges). To address the scalability limitations of bipartite quantum protocols, quantum network infrastructure must therefore become dynamic, scalable and protocol-aware, supporting a growing number of users, protocols and application requirements~\cite{Fitzke2022ScalableQKDNetwork,Cao2022EvolutionQKDNetworks,Martin2024MadQCI}. 
It is essential to incorporate entanglement-based networking, as only entanglement can support linked quantum computers. Further, it can be generated as a shared resource and dynamically distributed across different users to support advanced quantum communication applications beyond cryptographic use cases.


\begin{figure*}[!h]
    \centering
    \includegraphics[width=\linewidth,trim={0 0 0 0 },clip]
    {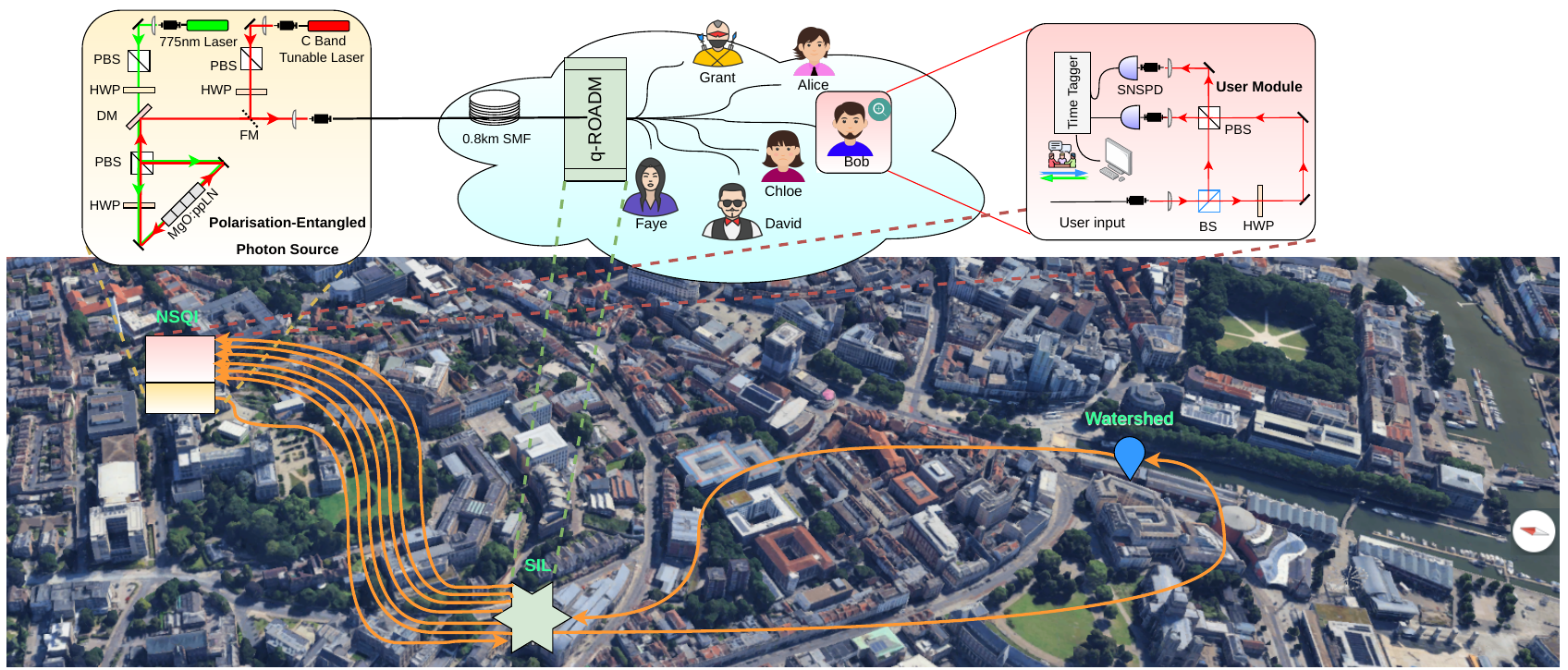}
    \caption{A 6-user entanglement-based quantum communication network architecture enabled by a q-ROADM. A type-0 SPDC broadband polarisation entangled photon source is connected to a q-ROADM via 0.8~km campus fibre between SIL and NSQI, supporting scalable, dynamic entanglement distribution. The 6 user modules are also deployed at NSQI connecting to the q-ROADM via the loop-back of another 0.8 km campus link for user A, B,C,F,G, and Bristol metropolitan fibre looping back from Watershed (user D). At each user, there is a polarisation analysis module, two SNSPDs, a time tagger and a computer to exchange the classical information with other users. 
    }
    \label{fig:Experiment_setup}
\end{figure*}

At the physical layer, a star topology is achieved by employing a central entangled photon source to distribute correlated photon pairs to different users.
Many entanglement-based quantum networks have primarily relied on passive distribution architectures, where entangled photon pairs are distributed through fixed optical paths using wavelength-division multiplexing (WDM) and/or beam splitters~\cite{qi2010feasibility,frohlich2013quantum,wengerowsky2018entanglement,joshi2020trusted,aktas2016entanglement,liu202240,qi202115,ciurana2014entanglement,fan2025quantum}. In particular, a polarisation-entangled source has been used to support a fixed-topology four-user quantum network~\cite{wengerowsky2018entanglement}. This concept was extended in~\cite{joshi2020trusted}, where dense WDM (DWDM) and beam splitters (BSs) were used to realise an eight-user entanglement-based quantum communication network. Beam splitter-based entanglement distribution has also been explored as a route to interconnecting a larger number of users~\cite{liu202240}, while entanglement-based quantum secure direct communication has been demonstrated for 15 users~\cite{qi202115}. However, passive entanglement distribution architectures only provide fixed connectivity, limiting dynamic reconfiguration capability. Such architectures also over provision some links to the detriment of overall network performance due to the extra noise they introduce.

To overcome the limitation, active reconfigurable entanglement distribution has been investigated using path and wavelength switching for flexible network control~\cite{lingaraju2021adaptive,zhu2019toward,appas2021flexible,laudenbach2020flexible,chang2016experimental,herbauts2013demonstration,Clark2025CoexistenceHCF, jiang2025entanglement}. Active optical switches have enabled flexible allocation of entanglement resources~\cite{laudenbach2020flexible,chang2016experimental,herbauts2013demonstration}, while wavelength-selective switches (WSSs) have been used for adaptive bandwidth management in entanglement distribution~\cite{zhu2019toward,lingaraju2021adaptive}. The WDM-based approaches using an on-chip entanglement source were demonstrated in~\cite{appas2021flexible,jiang2025entanglement}, connecting users in laboratory fibre or field-deployed fibre with different configurations. In~\cite{fan2025quantum, liu2024reconfigurable, huang2026quantum}, the authors demonstrated flexible entanglement distribution networks with dual or multiple pump wavelengths, achieving network programmability by tuning the pump wavelengths for Spontaneous Parametric Down Conversion (SPDC)-based sources and the spontaneous four-wave mixing source.
Although above demonstrations represent important steps towards flexible quantum networking, existing approaches have not yet provided a metropolitan-scale, multi-user entanglement distribution network that simultaneously supports dynamic topology control, long-term operation, multiple network configurations and protocol-level demonstrations over deployed fibre.

In this paper, we present a metropolitan-scale quantum communication network that supports dynamic polarisation-entanglement distribution to up to six users over the University of Bristol campus and metropolitan fibre links in Bristol. The network is enabled by a quantum reconfigurable optical add-drop multiplexer (q-ROADM), which combines wavelength multiplexing, optical switching and polarisation control to realise programmable allocation of correlated wavelength-channel pairs from a broadband entangled-photon source. This allows the same physical infrastructure to establish different quantum-network topologies, including full-mesh, partial-mesh and sliced sub-network configurations, according to user demand, link condition and protocol requirement. We first demonstrate a six-user fully-connected network supporting QKD, showcasing its $\approx$7-day stability over deployed fibre with secret key rate (SKR) ranging between 4.8 bps and 190 bps. We then compare full-mesh operation with dynamically switched partial-mesh configurations, showing how different entanglement-allocation strategies affect the accumulated secret keys under different source and pump-power conditions. Furthermore, we demonstrate quantum network slicing, where the infrastructure is reconfigured into independent sub-networks with optional interconnection links for service separation/connection. Finally, we show how the dynamic entanglement distribution network can support and optimise the implementation of the Secure Inaugural Authentication-Transfer (SIAT) protocol combined with flooding-based multi-path key concatenation~\cite{solomons2022scalable}, saving up to 2587~s for executing the SIAT protocol. These results show that q-ROADM-enabled entanglement distribution provides a practical route towards multi-user and multi-protocol quantum networking over deployed optical fibre infrastructure.

\section{Experimental Setup}

The schematic of the experiment is illustrated in~\Cref{fig:Experiment_setup}, consisting mainly of three parts: 1) a broadband polarisation-entangled photon source, 2) q-ROADM to enable flexible and on-demand entanglement distribution, and 3) users equipped with measurement modules to perform entanglement correlation. The entangled photon source is located within the Centre for Nanoscience \& Quantum Information (NSQI) building, connecting to the q-ROADM at Smart Internet Lab (SIL) in Merchants Venturers Building via 0.8~km campus fibre. The q-ROADM supports arbitrary shapes of entanglement distribution, which can establish quantum links between any combination of users. In this work, users were deployed at NSQI connecting to the q-ROADM either via direct loopback 0.8~km campus links or via an extra 4.8~km loopback metropolitan fibre via Watershed between the q-ROADM and user David.

As shown in~\Cref{fig:Experiment_setup}, a continuous wave laser operating at 775.06~nm (shown as green) emits light passing through a polarisation beam splitter (PBS) and a half-wave plate (HWP),  pumping the Sagnac loop with diagonally polarised light through a dichroic mirror (DM) and a dual-band PBS, where 50\% light propagates clockwise, defined as of vertically (\textit{V}) polarisation, and 50\% of the light propagates counterclockwise, defined as horizontally (\textit{H}) polarisation. A HWP inside the Sagnac loop is rotated at $45^{\circ}$ to convert \textit{V} light to \textit{H} light and vice versa. The light propagating along both directions is focused into the centre of a type-0 Magnesium-Oxide-Doped Periodical Poled Lithium Niobate (MgO:PPLN) crystal. The photons generated in either direction interfere with the vacuum state of the other in the PBS, therefore generating $\ket{\Phi^+} = \frac{1}{\sqrt{2}}(\ket{HH} + \ket{VV})$ state. Following the energy conservation law, the wavelengths of non-degenerate photon pairs are symmetric to 1550.12~nm. DM reflects entangled photons by isolating them from the 775.06~nm pump light, which are further coupled into the fibre network. Due to birefringence, the polarisation of the photons changes and fluctuates when transmitting over the fibre. Therefore, we have to stabilise the polarisation to ensure all the users measure with the agreed polarisation basis. We call this process fibre polarisation neutralisation, which is further detailed in \cref{sec:polarisation_neutralisation} of the Supplement.

\begin{figure}[!h]
    \centering
    \includegraphics[width=\linewidth,
    ]{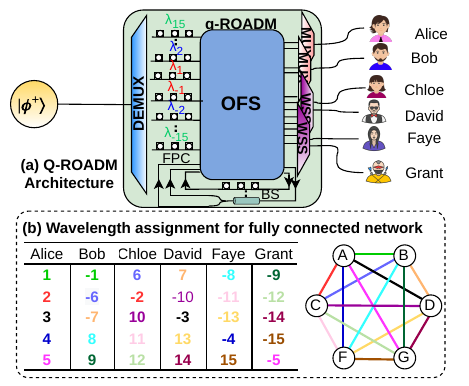}
    \caption{(a) q-ROADM architecture for dynamic entanglement distribution to multiple users. (B) Wavelength assignment for the six-user full-mesh configuration, where each table entry denotes the channel $\lambda_i$ assigned to the corresponding link.}
    \label{fig:Q_ROADM}
\end{figure}

\begin{table}[t]
\centering
\setlength{\tabcolsep}{4pt}        
\renewcommand{\arraystretch}{0.92}  
\caption{Fibre distance and channel loss from source to each user.}
\label{tab:user_loss}
\small
\begin{tabular}{@{}lcc@{}}
\toprule
\textbf{User} & \textbf{Distance (km)} & \textbf{Loss (dB)} \\
\midrule
Alice & 1.6 & 8.9--10.0 \\
Bob   & 1.6 & 8.1--11.1 \\
Chloe & 1.6 & 10.6--13.0 \\
David & 5.6 & 14.1--15.1 \\
Faye  & 1.6 & 10.6--11.9 \\
Grant & 1.6 & 11.0--11.9 \\
\bottomrule
\end{tabular}
\end{table}

The architecture of the q-ROADM is illustrated in \Cref{fig:Q_ROADM}a, where a demultiplexer (DEMUX) divides the spectrum of the broadband entangled photon source into 30 slices according to ITU-T Rec. G.694.1 100~GHz DWDM standard. The central wavelength $\lambda_{0}$ of the DEMUX is ITU-T channel 34 with the wavelength of 1550.12~nm, corresponding to the down-converted spectral centre of the source. We denote the rest wavelengths as $\lambda_{\pm i}$ or $\pm i$, where $i$ is an integer in the range of $[-15,15]$. The wavelengths from the non-degenerate SPDC process are equally spaced and can be abstracted in the form of wavelength pairs [$\lambda_{i}$, $\lambda_{-i}$], where photons in $\lambda_{i}$ are polarisation entangled with those in $\lambda_{-i}$. 
Each 100~GHz channel is connected to a fibre polarisation controller (FPC) to stabilise the polarisation, as discussed in \cref{sec:polarisation_neutralisation} of Supplement. The FPCs are then connected to a $192 \times 192$ optical fibre switch (OFS, Polatis from HUBER+SUHNER), in which any cross-connection can be programmed. The wavelength-selective switch (WSS) and multiplexer (MUX) combine different wavelength channels into a single fibre link connected to the individual user. In the experiment, we use two $1 \times 16$ MUXs connecting to Alice (A) and Bob (B), respectively, and two $4 \times 16$ WSSs (Finisar WaveShaper 16000S), one connected to Chloe (C) and David (D) and the other one supporting Faye (F) and Grant (G). By jointly programming OFS and WSS, the q-ROADM can distribute different entanglement pairs to different users simultaneously. \Cref{fig:Q_ROADM}b shows a wavelength assignment scheme for a 6-user full mesh quantum communication network configuration. The assignment scheme utilises all the 15 pairs of entangled wavelengths to support 15 quantum links. More details of the q-ROADM reconfigurability is described in \cref{sec:qroadm_supplement} of Supplement.


\begin{figure}[h]
\centering
\includegraphics[width=\linewidth]{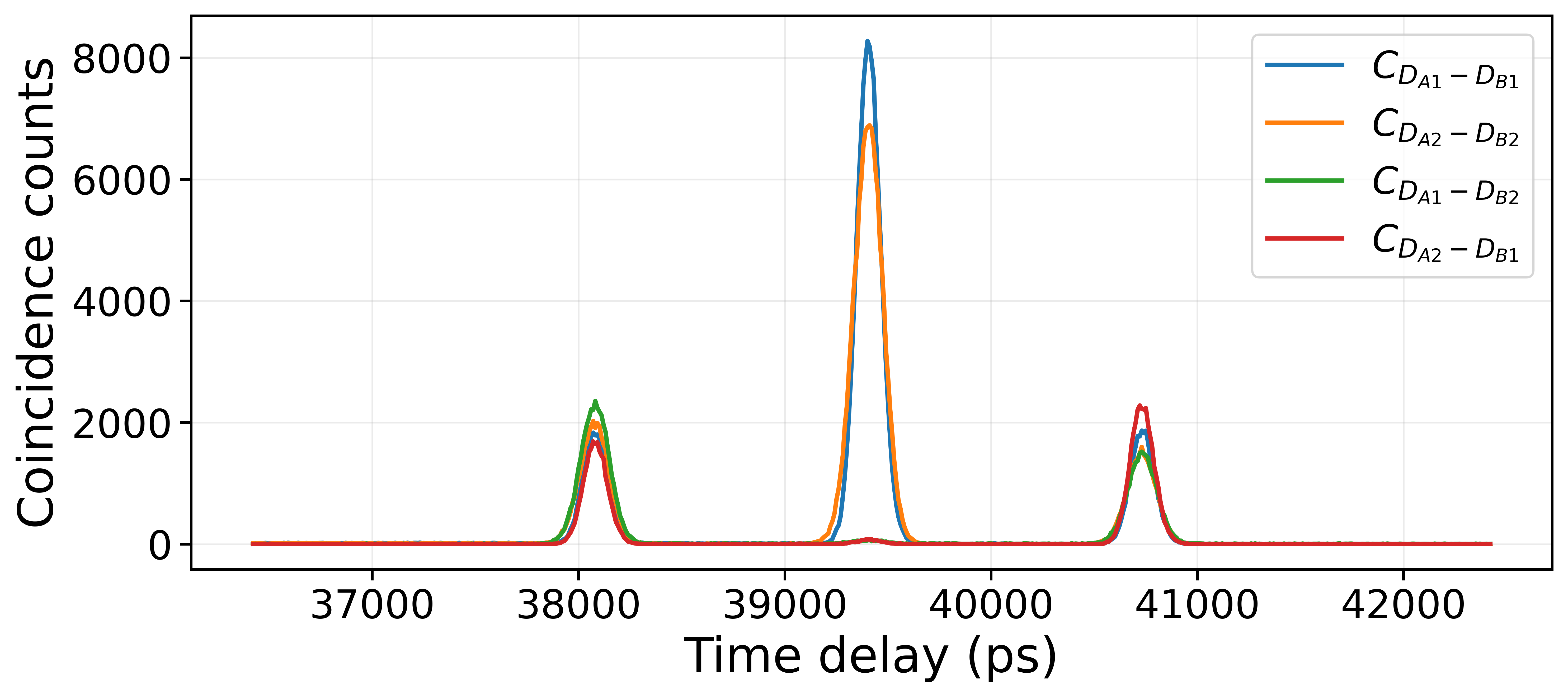}
\caption{Temporal correlation histogram between Alice and Bob when the network is configured to support only the A--B link, with 10-minute integration time and binwidth of 10~ps.}
\label{fig:g2_diagram}
\end{figure}

Each user, shown on the right of \Cref{fig:Experiment_setup}, consists of a polarisation analysis module (PAM) and two superconducting nanowire single-photon detectors (SNSPDs). The fibre input is collimated in free space and sent to a beam splitter (BS), which passively selects the measurement basis with photons travelling into either a short or long optical path. The short path performs an HV-basis measurement using the PBS, while the long path includes an HWP set to $22.5^{\circ}$ to enable a DA-basis measurement using the same PBS and detector pair.
The two PAM outputs are coupled into fibre and connected to two SNSPDs. For six users, the system therefore uses 12 SNSPD channels connected to a Swabian Time Tagger Ultra. The measured coupling efficiencies of the PAMs range from 52.8\% to 76.1\%, with the imbalance between the short and long paths controlled within 4\%. 
The fibre distance and loss from the source to each user via the q-ROADM are summarised in \Cref{tab:user_loss}. The loss varies due to different fibres, their lengths, and q-ROADM with different wavelength assignment for different connections, as detailed in Fig. \ref{fig:Q_ROADM} (b).

\section{Results}
\label{Sec:Results}

In this work, we use the SKR as the main metric to evaluate the performance of the dynamic entanglement network supporting the BBM92 protocol~\cite{bennett1992quantum}. This allows us to capture both the capacity of each link and its quality, including the impact of loss, accidental coincidences and detector timing jitter. Experimentally, we investigate two operating conditions: a favourable case with an entanglement source heralding efficiency (HE) of approximately 15\% and user-side SNSPD jitter below 100~ps, and a more challenging case with approximately 3\% HE and detector jitter of 300--350~ps.
With the q-ROADM configured to distribute the appropriate wavelength pairs, each user exchanges time-tagging information with its communication partner to execute the BBM92 protocol. Temporal correlations are then obtained by combining the local time tags with those received from the remote user. An example 10-minute temporal cross-correlation histogram, $g^{(2)}$, for the Alice--Bob link under the favourable operating condition is shown in \Cref{fig:g2_diagram}, with a bin width of 10~ps. The four curves correspond to the coincidence histograms $C_{D_{A1}-D_{B1}}$, $C_{D_{A2}-D_{B1}}$, $C_{D_{A1}-D_{B2}}$ and $C_{D_{A2}-D_{B2}}$, where $C_{D_{Ai}-D_{Bj}}$ denotes the coincidence counts between detector $i$ of Alice and detector $j$ of Bob, with $i,j\in{1,2}$. Due to the short and long path architecture of the user modules, the central peaks correspond to events where both users measure in the same polarisation basis. The two side peaks arise from events where one photon takes the short path, and the other takes the long path or vice versa.

\begin{figure}[h]
    \centering
    \includegraphics[width=\linewidth]{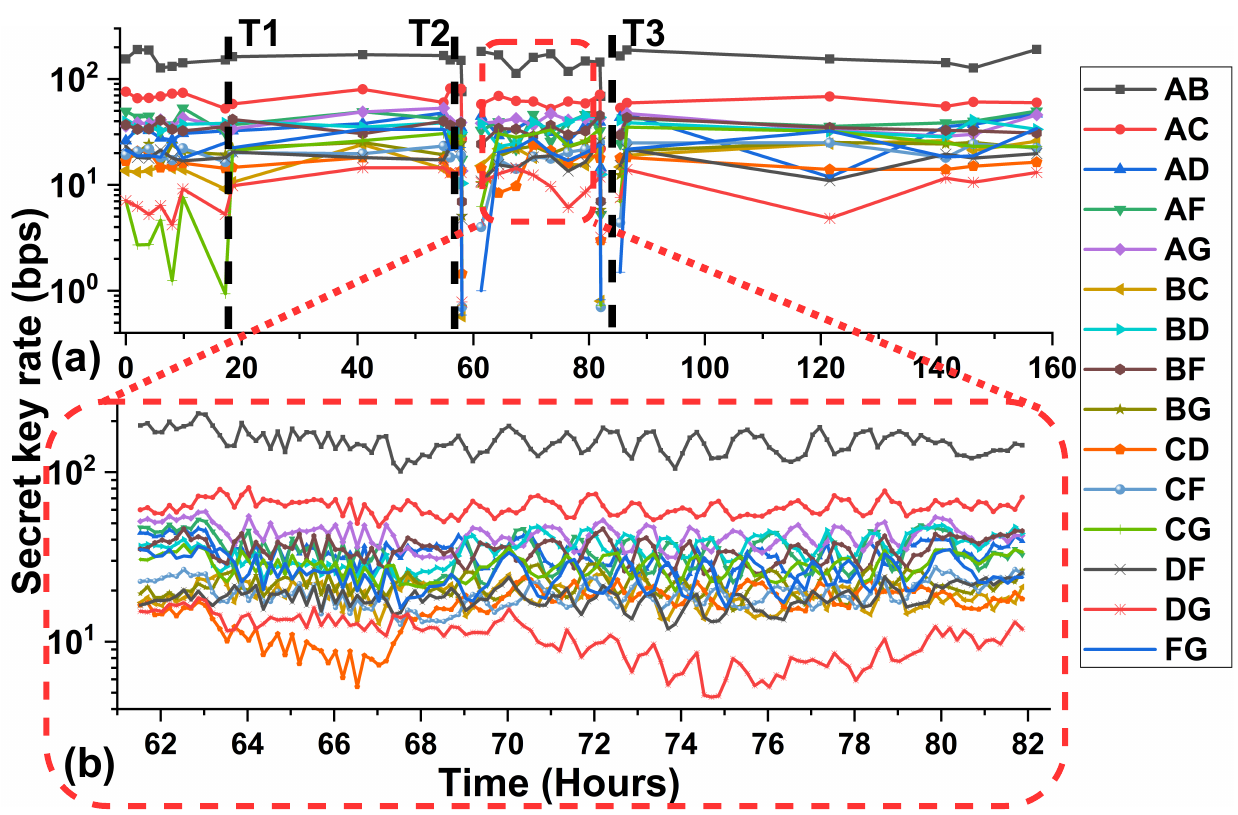}
    \caption{SKR of all 15 links in the six-user full-mesh quantum network. (a) SKR over $\approx$7 days; (b) zoomed-in SKR over $\approx$20 hours.}
    \label{fig:Long_term_monitoring}
\end{figure}

\begin{figure}[h]
    \centering
    \includegraphics[width=\linewidth]{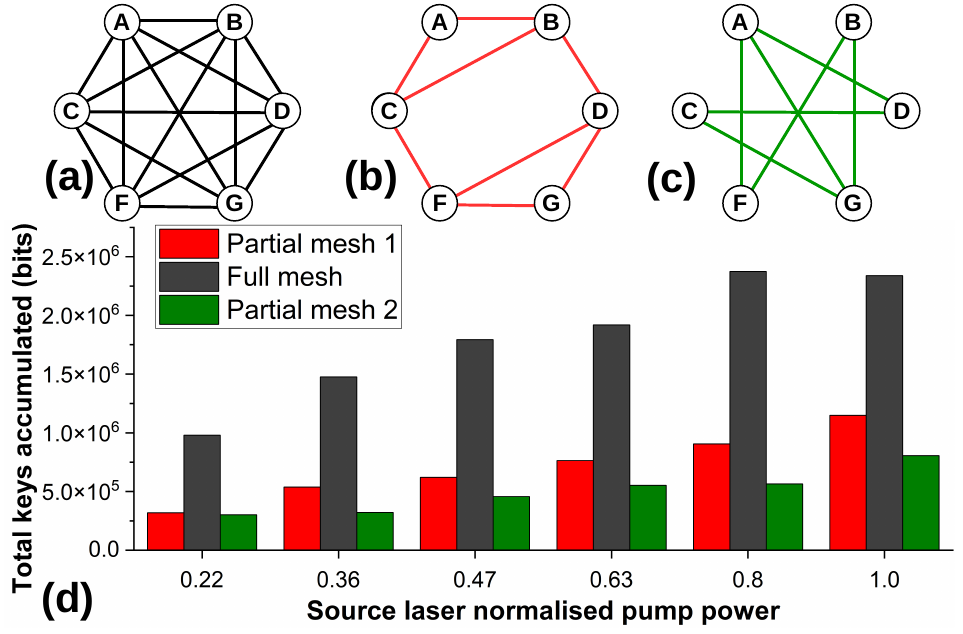}
    \caption{Comparison between full-mesh and time-shared partial-mesh configurations under different normalised source pump powers. (a) Six-user full-mesh topology. (b,c) Two complementary partial-mesh topologies. (d) Total accumulated secret keys over 40 minutes for the full-mesh configuration, compared with two 20-minute partial-mesh configurations under the favourable network operating condition.}
    \label{fig:FM_PM_Keys_Power}
\end{figure}

\begin{figure*}[!h]
    \centering
    \includegraphics[width=\textwidth,height=0.72\textheight,keepaspectratio]{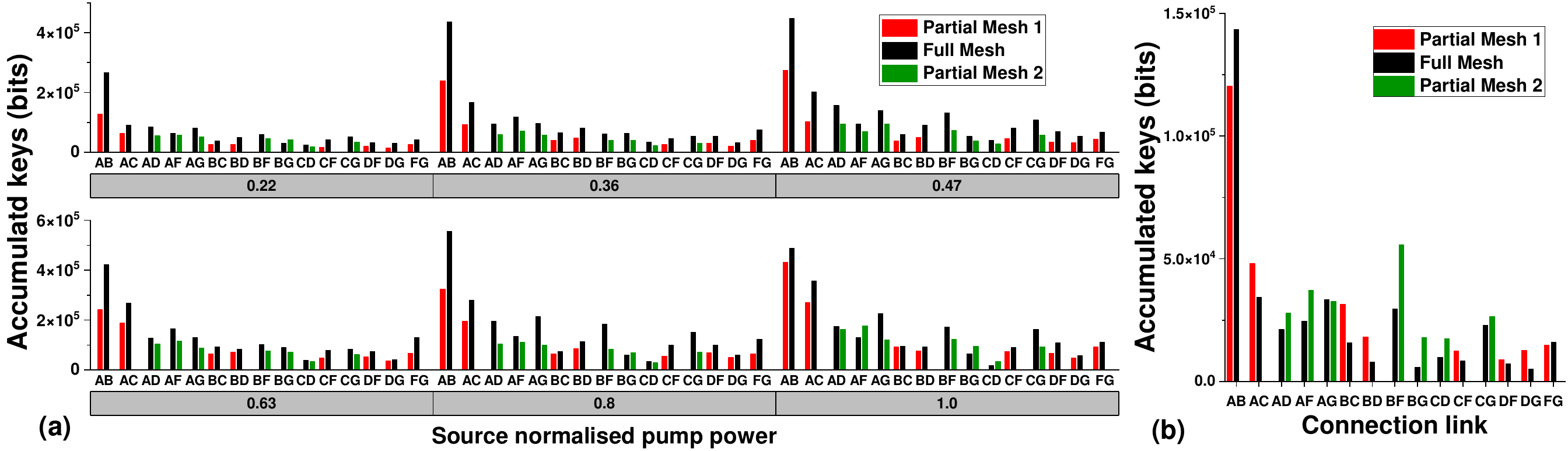}
    \caption{Accumulated secret keys for individual quantum links under full-mesh and time-shared partial-mesh configurations. (a) Link-level comparison under the favourable operating condition. (b) Link-level comparison under the challenging operating condition with approximately 3\% source heralding efficiency and 300--350~ps detector jitter.}
    \label{fig:FM_PM_links_power}
\end{figure*}

Given a fixed coincidence window, the total number of coincidences is obtained by summing all detector-pair combinations:
\begin{equation}
C_{\mathrm{total}} =
C_{D_{A1}-D_{B1}} +
C_{D_{A2}-D_{B1}} +
C_{D_{A1}-D_{B2}} +
C_{D_{A2}-D_{B2}} .
\label{eq:coincidence}
\end{equation}
For the Bell state considered here, the events $C_{D_{A1}-D_{B1}}$ and $C_{D_{A2}-D_{B2}}$ are assigned as correct bits, while $C_{D_{A2}-D_{B1}}$ and $C_{D_{A1}-D_{B2}}$ contribute to the error counts. The QBER of the Alice--Bob link is therefore calculated as
\begin{equation}
Q_{\mathrm{A-B}} =
\frac{
C_{D_{A2}-D_{B1}} +
C_{D_{A1}-D_{B2}}
}{
C_{\mathrm{total}}
}.
\label{eq:qber}
\end{equation}
The sifted key length, $N_{\mathrm{sifted}}$, is obtained from the coincidences within the selected basis by assigning binary digits to the correlated detector outcomes $C_{D_{A1} - D_{B1}}$ and $C_{D_{A2} - D_{B2}}$. After distillation, including error correction and privacy amplification, the final secret key length is lower bounded in the asymptotic limit by:
\begin{equation}
N_{\mathrm{key}} \geq
N_{\mathrm{sifted}}
\left[
1 - f_{\mathrm{EC}} H_2(Q_{\mathrm{A-B}})
  - H_2(Q_{\mathrm{A-B}})
\right],
\label{eq:skr}
\end{equation}
where $f_{\mathrm{EC}}$ is the error-correction efficiency and $H_2(p)$ is the binary entropy function,
\begin{equation}
H_2(p) = -p\log_2(p) - (1-p)\log_2(1-p).
\end{equation}
Here, we assume that the phase error rate is equal to the measured bit error rate. It is worth noting that although Alice and Bob are used as the example in \Cref{eq:qber,eq:skr}, the same analysis is applied to all other links or user pairs in the network.

\subsection{Stable entanglement distribution for a fully connected six-user quantum network}

To evaluate the stability of the proposed q-ROADM architecture, the network was configured as a six-user full-mesh setting using the wavelength-pair assignment shown in \Cref{fig:Q_ROADM}b. The experiment was performed under the favourable operating condition described earlier. The measured asymptotic secret key rates of all 15 quantum links are shown in \Cref{fig:Long_term_monitoring}. Each data point represents the average SKR over a 10-minute integration time. As shown in \Cref{fig:Long_term_monitoring}a, stable entanglement distribution was maintained over 157.3~hours. The best-performing link was A--B, with an average SKR of 153.2~bps, due to its lower end-to-end loss. In contrast, the D--G link with the highest loss achieved an average SKR of 9.8~bps. At $T1=18.43$~h, the fibre polarisation of all users was re-neutralised after the SKR of several links began to degrade. The interruptions at $T2=58.03$~h and $T3=82.03$~h were caused by regular cycling of the SNSPD system.

A zoomed-in view of the continuous SKR data between 61.37~h and 81.87~h is shown in \Cref{fig:Long_term_monitoring}b. Along the full measurement period, the quantum network was also reconfigured for other experimental scenarios between $T1$ and $T2$, and again after $T3$. A power outage lasting more than 24~hours occurred a few hours after $T3$. The results confirm that the q-ROADM-based quantum network can maintain stable multi-user entanglement distribution over deployed fibre for more than 140~hours after polarisation re-neutralisation, even after network reconfiguration and power interruption.

\subsection{Full-mesh and partial-mesh entanglement distribution}

A QKD network can accumulate secret keys between user pairs and consume them later. Therefore, not all links necessarily need to remain active continuously, provided that sufficient keys have already been generated and safely stored. Here, we investigate dynamic entanglement allocation,  
and compare the six-user full-mesh configuration in \Cref{fig:FM_PM_Keys_Power}a with two complementary partial-mesh configurations shown in \Cref{fig:FM_PM_Keys_Power}b,c. The two partial meshes do not contain overlapping quantum links and, when combined, provide the same connectivity as the full-mesh network. For comparison, the full-mesh configuration is operated for 40~minutes, while each partial-mesh configuration is operated for 20~minutes, giving the same total measurement time.
The total accumulated secret keys under the favourable network condition with various source pump power are shown in \Cref{fig:FM_PM_Keys_Power}d. Across the investigated pump-power range, the full-mesh configuration generates more total keys than the two time-shared partial-mesh configurations. Increasing the pump power also increases both the coincidence rate and the accidental rate, which raises the QBER simultaneously. Depending on the network condition, this can increase or reduce the final SKR. As a result, the total key generation in the full-mesh configuration decreases at the highest pump power compared with 0.8 normalised power. For the partial mesh scheme, the optimum power is higher than that of the investigated schemes, with fewer accidentals due to fewer assigned channels compared to the full mesh scheme.

The link-level accumulated keys are shown in \Cref{fig:FM_PM_links_power}a for the favourable condition (the source HE is 15\%, and SNSPD jitter < 100 ps). Most links in the full-mesh configuration generate more keys than in the time-shared partial-mesh configurations. However, several links show reduced key generation at the highest pump power, consistent with the increase in QBER caused by accidentals. In contrast, the partial-mesh configurations continue to benefit from increasing pump power over the investigated range.
Under the challenging network condition with approximately 3\% source heralding efficiency and detector jitter of 300--350~ps, the coincidence peaks are broadened, and the ratio of true coincidences to accidental decreases. In this regime, the time-shared partial-mesh configurations can outperform the full-mesh configuration for many links, as shown in \Cref{fig:FM_PM_links_power}b, because reducing the number of simultaneous wavelength channels lowers the accidentals and improves the extractable key rate. These results show that the optimum entanglement-allocation strategy depends strongly on source brightness, heralding efficiency, detector timing jitter and link loss. The q-ROADM, therefore, can provide a useful mechanism for adapting the network topology to the operating condition and quantum service requirement.

\begin{figure*}[!h]
    \centering
    \includegraphics[width=\linewidth]{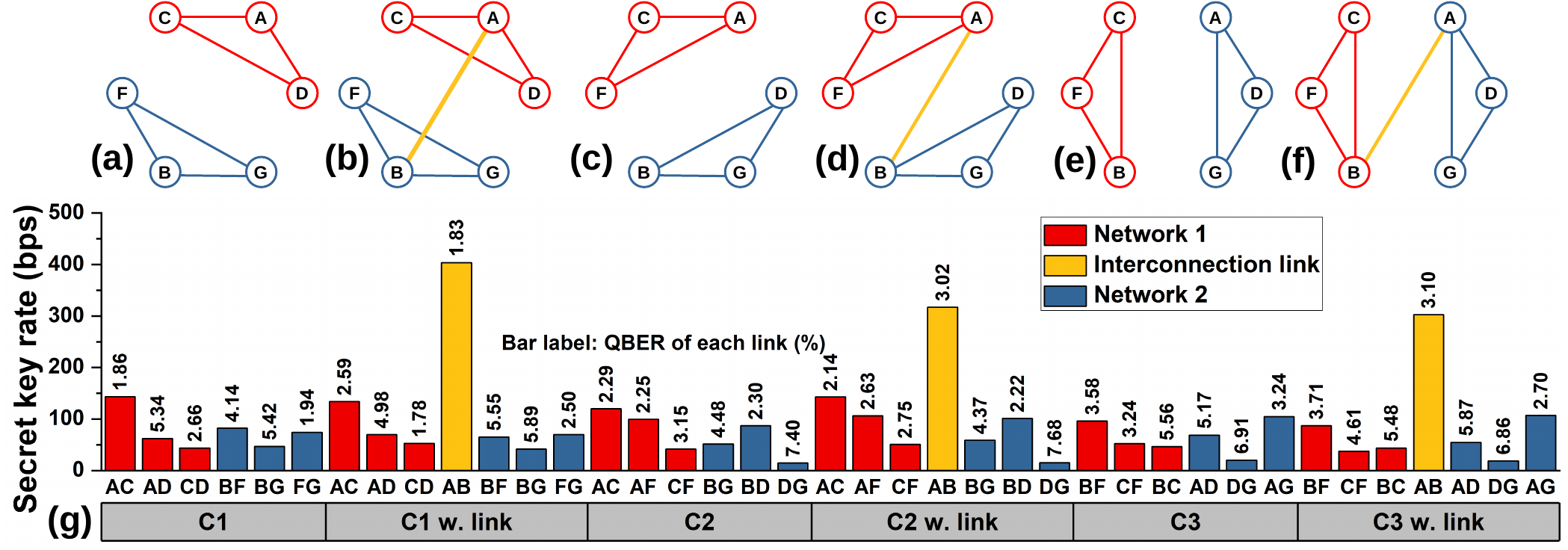}
    \caption{Quantum network slicing in the six-user entanglement network. (a,c,e) Three slicing cases, C1--C3, where the physical network is partitioned into two independent three-user sub-networks. (b,d,f) Corresponding configurations with an additional A--B interconnection link between the two sub-networks. Red and blue denote Network~1 and Network~2, respectively, while yellow denotes the interconnection link. (g) SKR and QBER performance for each sliced-network configuration. Bar heights show SKR, and labels indicate QBER in percentage. The results show that the interconnection link can be introduced without significantly degrading the performance of the sliced sub-networks.}
    \label{fig:2Nets}
\end{figure*}

\subsection{Quantum network slicing}

The flexible entanglement distribution enabled by the q-ROADM allows the same physical quantum network to be partitioned into multiple logical sub-networks. This capability is analogous to network slicing in classical networks, where a shared infrastructure can be divided into logically separated services with different users, connectivity requirements. In the context of quantum networking, such slicing is particularly useful because entanglement resources are limited and must be allocated according to user demand, link quality and protocol requirement. A dynamically reconfigurable quantum network can therefore support multiple independent quantum services while avoiding the need to duplicate the underlying physical infrastructure.

Here, we demonstrate quantum network slicing by partitioning the six-user entanglement network into two independent three-user sub-networks. Three slicing configurations are shown in \Cref{fig:2Nets}a,c,e, denoted as Case 1 (C1), Case 2 (C2) and Case 3 (C3), respectively. In each case, the red links form Network~1 and the blue links form Network~2. The wavelength pairs are assigned by the q-ROADM such that the two sub-networks can operate independently, with no quantum links between them.
In addition to isolated sub-networks, the q-ROADM can selectively establish an interconnection link between two sliced sub-networks. The corresponding configurations with an additional A--B interconnection link are shown in \Cref{fig:2Nets}b,d,f, where the interconnection link is highlighted in yellow. Such an interconnection can support communication between otherwise independent sub-networks, enabling temporary service federation or controlled information exchange. The link can be added or removed by reconfiguring the q-ROADM, without physically modifying the network. The SKR and QBER of each link are shown in \Cref{fig:2Nets}g, where the bar height represents the SKR and the label above each bar gives the corresponding QBER. The results show that adding the A--B interconnection link has no significant impact on the performance of the quantum links within the two sliced sub-networks. The variations in SKR and QBER are mainly attributed to performance fluctuations over different measurement periods, similar to the long-term monitoring results in \Cref{fig:Long_term_monitoring}.

\subsection{q-ROADM-enabled SIAT and flooding protocol}
\begin{figure}[h]
\centering
\includegraphics[width=\linewidth]{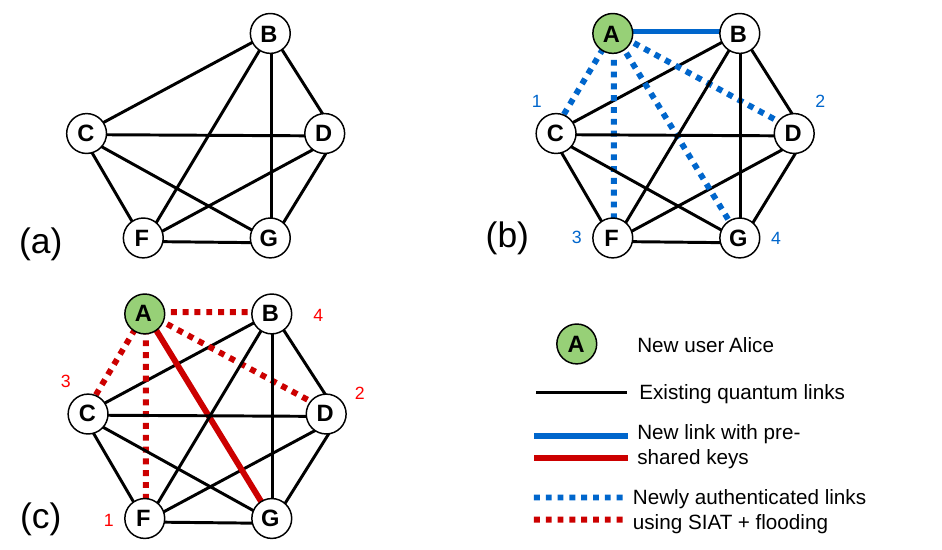}
\caption{SIAT and flooding protocol for authenticating a new user, Alice, into an existing five-user entanglement network. (a) Initial five-user fully connected network. (b) Strategy~1: A pre-shares authentication keys with B and is then authenticated with users C, D, F and G. (c) Strategy~2: A pre-shares authentication keys with G and is then authenticated with users F, D, C and B. Solid black lines denote existing quantum links, solid coloured lines denote the initial pre-shared-key link, dotted coloured lines denote links authenticated using SIAT and flooding and the number by the user is the authenticating order.}
\label{fig:SIAT_diagram}
\end{figure}

A quantum network should distribute entanglement resources to meet the quantum network protocols' needs. Here, we demonstrate how the q-ROADM-enabled dynamic entanglement network can support the Secure Inaugural Authentication-Transfer (SIAT) protocol combined with flooding-based multi-path key establishment~\cite{solomons2022scalable}. The SIAT protocol is used for authentication transfer when a new user joins an existing quantum network. Before the new user can establish quantum communication with other users, its identity must be authenticated. In a fully connected network with \textit{N} existing users, this can require up-to \textit{N} authenticated connections. Therefore, scalable authentication-transfer mechanisms using entanglement links are required.
As shown in \Cref{fig:SIAT_diagram}a, we consider an existing five-user fully connected entanglement network consisting of B, C, D, F and G. A new user, Alice, is added to the network through the q-ROADM via an additional MUX connected to the OFS. The q-ROADM then dynamically configures only the quantum links required at each authentication step.

In SIAT, the new user initially shares an authentication key with at least one trusted user in the network. This trusted user can then transfer authentication to another user. For example, in Strategy~1, A first shares authentication keys with B, who acts as the trusted intermediary for authenticating A with C. Once A--C link has generated sufficient keys through their direct entanglement link, trust in Bob is no longer needed for future A--C authentication. Subsequent rounds can be authenticated using information generated by the previous round of the direct A--C link. This procedure is repeated until A is authenticated with all target users. When multiple already-authenticated paths are available, flooding can be used to transfer authentication through non-overlapping paths. In the security-enhancing mode considered here, keys generated along different paths are XORed, reducing reliance on any single trusted intermediate node.

We compare two authentication strategies, shown in \Cref{fig:SIAT_diagram}b and \Cref{fig:SIAT_diagram}c. In Strategy~1, Alice initially pre-shares authentication keys with Bob and is authenticated with the remaining users in the order C, D, F and G. In Strategy~2, Alice initially pre-shares authentication keys with Grant and is authenticated in the reverse order, F, D, C and B. The required authentication key length is calculated using Wegman--Carter authentication~\cite{wegman1981new}, and the timing analysis uses the measured SKRs of the dynamically configured quantum links.

\begin{figure*}[h]
\centering
\includegraphics[width=\linewidth]{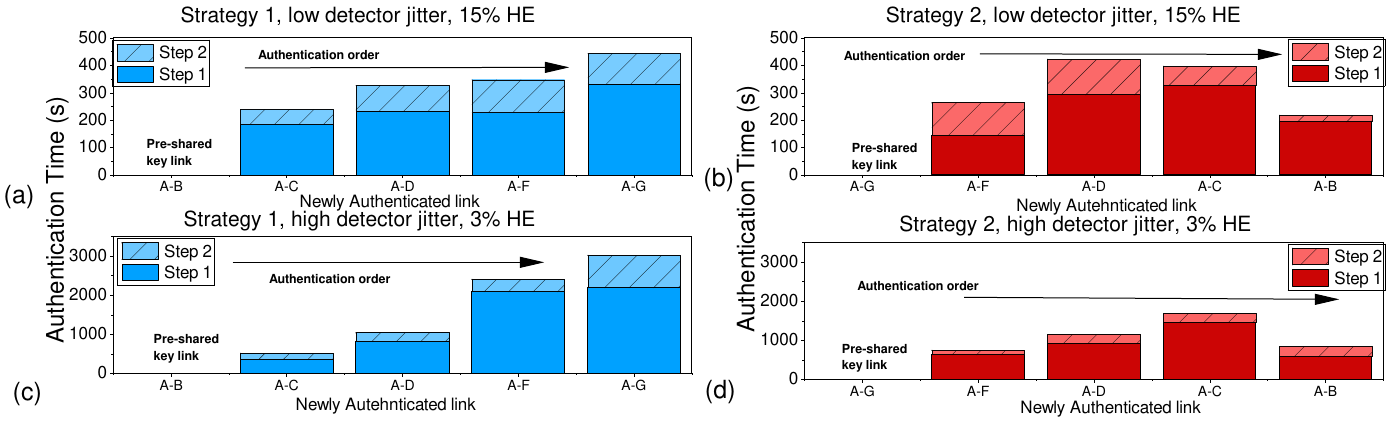}
\caption{Time required to authenticate new user Alice into the five-user network using SIAT and flooding. Each bar is divided into Step~1, corresponding to flooding-based authentication transfer, and Step~2, corresponding to direct entanglement-based key generation for subsequent authentication. (a) Strategy~1 under the favourable condition with low detector jitter and approximately 15\% source heralding efficiency. (b) Strategy~2 under the favourable condition. (c) Strategy~1 under the challenging condition with 300--350~ps detector jitter and approximately 3\% source heralding efficiency. (d) Strategy~2 under the challenging condition.}
\label{fig:SIAT_results}
\end{figure*}

The SIAT execution time depends on the SKR of the relevant quantum links, which is affected by link loss, detector efficiency, timing jitter, source heralding efficiency and accidental coincidences. We therefore evaluate the two strategies under two representative operating conditions: a favourable condition and a challenging condition as specified at the beginning of the section \ref{Sec:Results}.
The results are shown in \Cref{fig:SIAT_results}. Under the favourable condition, Alice is authenticated with the full five-user network in 1356~s using Strategy~1 and 1298~s using Strategy~2. The two strategies therefore perform similarly when the quantum links have relatively high SKR. Under the challenging condition, the authentication order has a much stronger impact: Strategy~2 completes in 4371~s, while Strategy~1 needs 6958~s. This difference arises because the flooding-based transfer step is limited by the weakest selected path, and this bottleneck becomes more pronounced when the source heralding efficiency is low and the detector timing jitter is high. In addition, different authentication orders require different intermediate links and network configurations. These links can exhibit different QBER and SKR, making the overall SIAT time more sensitive to the authentication order under challenging network conditions.
These results show that by configuring only the links required for each SIAT step, the q-ROADM acts as a protocol-aware entanglement distribution layer, enabling authentication-transfer protocols to be implemented and optimised according to the measured link conditions. A detailed SIAT timing analysis is provided in \cref{sec:siat_time_analysis} of the supplement.

\section{Conclusion}

We have demonstrated a dynamic entanglement-based quantum communication network supporting multi-user and multi-protocol operation over deployed campus and metropolitan fibre. The network is enabled by a q-ROADM that combines DWDM channelisation, optical switching, wavelength multiplexing and polarisation control to distribute correlated wavelength-channel pairs from a broadband polarisation-entangled photon source. Using this architecture, we realised programmable entanglement distribution among six users, including full-mesh, partial-mesh and sliced sub-network configurations. The six-user full-mesh network supported all 15 pairwise quantum links and maintained stable secret-key-rate performance over more than 150~hours, confirming the feasibility of long-term dynamic entanglement distribution over deployed fibre infrastructure.
Beyond static key-generation demonstrations, we showed that the q-ROADM provides a protocol-aware entanglement distribution layer. By reconfiguring the same physical infrastructure, the network can: adapt between full-mesh and time-shared partial-mesh operation according to demand, entanglement source and detector conditions; support quantum network slicing with optional/additional interconnection links; and enable SIAT combined with flooding-based multi-path authentication for onboarding a new user. These results highlight that the optimum entanglement-allocation strategy depends on link loss, source heralding efficiency, detector jitter and service requirement. The demonstrated q-ROADM-enabled architecture therefore provides a practical route towards scalable, reconfigurable and service-oriented quantum networks capable of supporting multiple users and multiple quantum networking protocols over shared optical fibre infrastructure.

\section{Funding Acknowledgement}
We acknowledge the funding from the UK's Engineering and Physical Sciences Research Council projects - Quantum Communications Hub (EP/T001011/1), Integrated Quantum Networks Hub (EP/Z533208/1), Towards The Quantum Internet (EP/X039439/1) and Science and Technology Facilities Council DyMEND project (UKRI593).

\bibliographystyle{IEEEtranN}

\bibliography{Main}

\clearpage
\section*{Supplementary Material}

\setcounter{subsection}{0}
\renewcommand{\thesubsection}{S\arabic{subsection}}

\setcounter{figure}{0}
\renewcommand{\thefigure}{S\arabic{figure}}

\setcounter{table}{0}
\renewcommand{\thetable}{S\arabic{table}}

\setcounter{equation}{0}
\renewcommand{\theequation}{S\arabic{equation}}

\begin{figure*}[!h]
\centering
\includegraphics[width=\textwidth]{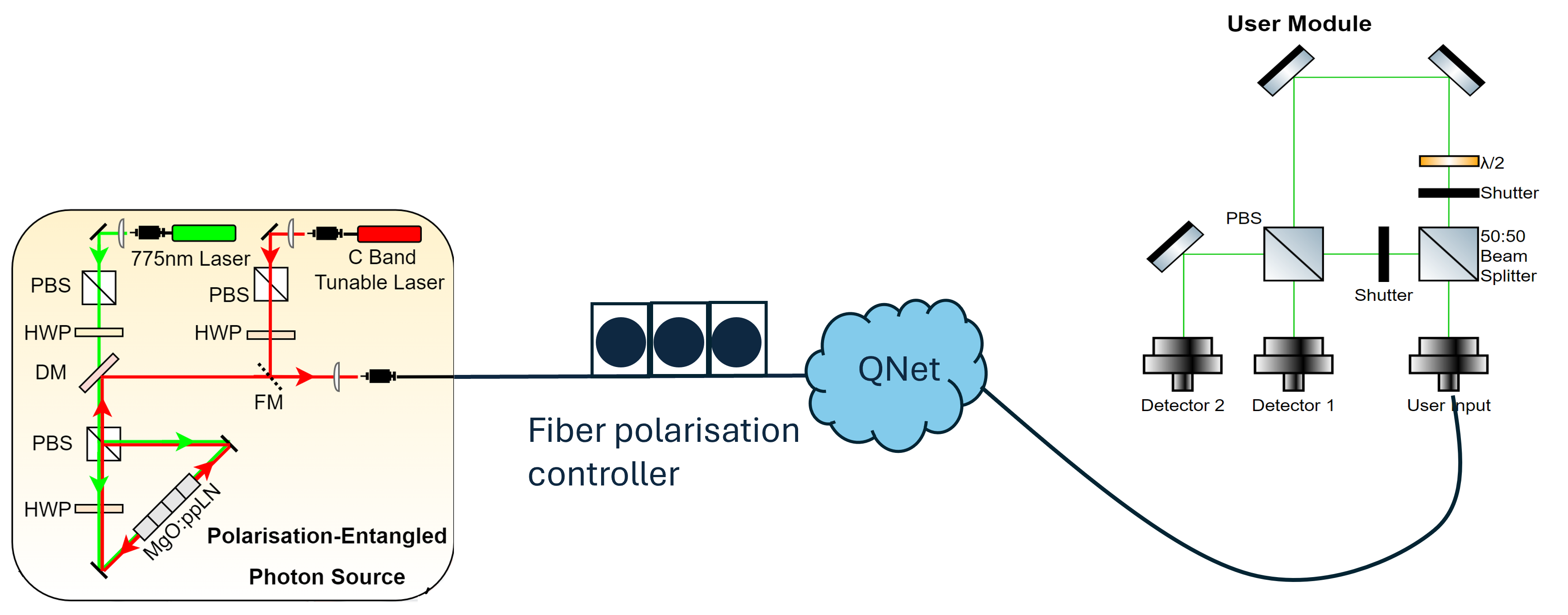}
\caption{Fibre polarisation neutralisation scheme for wavelength-multiplexed entanglement distribution. A tunable C-band laser provides predefined H- and D-polarised weak coherent reference states, which are injected into the same network path as the entangled photons. The FPCs inside the q-ROADM are adjusted so that the reference state is recovered at the selected user module, ensuring that all users measure in the agreed H/V and D/A bases.}
\label{fig:Neutralisation}
\end{figure*}

\subsection{Quantum Network Polarisation Neutralisation}
\label{sec:polarisation_neutralisation}







Polarisation neutralisation is required for each quantum communication link because fibre birefringence induces wavelength-dependent and time-varying polarisation rotations. Without compensation, the measurement bases at the users would no longer correspond to the agreed H/V and D/A bases used for entanglement correlation and BBM92 key generation. In this work, we neutralise the fibre links by ensuring that predefined reference polarisation states launched from the source are recovered at the user modules after transmission through the q-ROADM and the deployed fibre paths.

The neutralisation architecture is illustrated in \Cref{fig:Neutralisation}. A C-band tunable laser, shown as the red laser, is attenuated to the single-photon level and used as a wavelength-selectable polarisation reference. The laser output is coupled into free space and passed through a high-extinction-ratio polarisation beam splitter (PBS), using a Wollaston prism (WPY10, Thorlabs). One output of the PBS is defined as the horizontally (H) polarised reference state. A half-wave plate (HWP), mounted on a motorised rotation stage, is placed after the PBS. With the HWP fast axis aligned to the horizontal polarisation, the output remains H-polarised; rotating the HWP by $22.5^{\circ}$ converts the reference to the diagonal (D) polarisation state.

A motorised flip mirror (FM) allows the source to switch between two operation modes. In the quantum communication mode, the entangled photon pairs from the polarisation-entangled photon source are coupled into the network. In the neutralisation mode, the predefined H- or D-polarised weak coherent reference light is injected from free-space into the same fibre path and routed through the q-ROADM to the selected user. At each user module, two shutters control the measurement path. During the neutralisation process, for H-polarised reference light, only the short path is opened so that the H/V basis is measured. For D-polarised reference light, only the long path is opened so that the D/A basis is measured. The fibre polarisation controller (FPC) inside the q-ROADM is then adjusted such that the count rate at the expected detector is maximised while the count rate at the orthogonal-basis detector is minimised.

This procedure is repeated for each assigned wavelength channel used by the entanglement distribution scheme. After all required wavelengths and user paths have been neutralised, the flip mirror is switched back to block the reference laser and allow the entangled photon pairs in the quantum network. In this way, the q-ROADM can distribute different entangled wavelength pairs while maintaining a common polarisation reference across all users and links.

\subsection{q-ROADM Architecture and Wavelength Assignment}
\label{sec:qroadm_supplement}

\begin{figure*}[h]
\centering
\includegraphics[width=\linewidth]{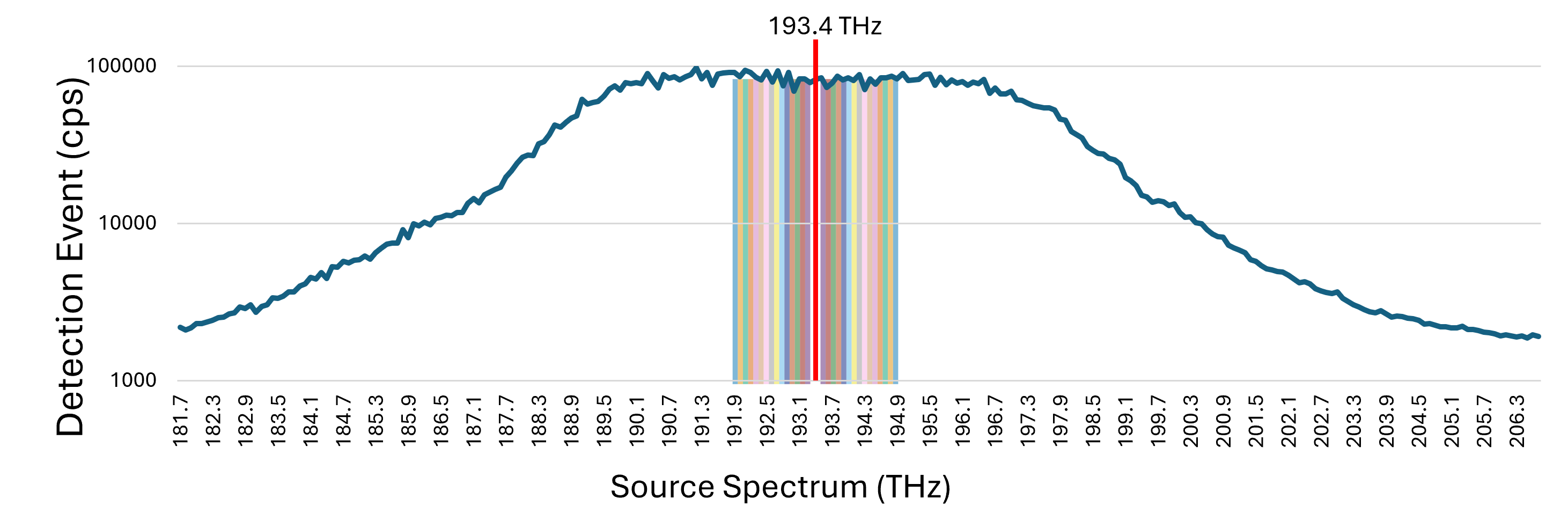}
\caption{Entanglement source characterisation showing the photon detection rate as a function of optical frequency. The central frequency is 193.4~THz, corresponding to 1550.12~nm, and the coloured bars indicate the 100~GHz DWDM channels used by the q-ROADM.}
\label{fig:source_spectrum}
\end{figure*}

The q-ROADM was designed to balance insertion loss, entanglement-distribution flexibility, and scalability towards a larger number of users. The broadband SPDC source has a full-width at half-maximum (FWHM) bandwidth of approximately 70~nm across the telecom S, C and L bands, centred at 1550.12~nm, corresponding to 193.4~THz. As shown in \Cref{fig:source_spectrum}, the central part of this spectrum is sliced into 30 non-overlapping 100~GHz DWDM channels using a passive demultiplexer (DEMUX) following the ITU-T grid. Due to energy conservation in the SPDC process, photons generated in channel $\lambda_i$ are correlated with photons in channel $\lambda_{-i}$. These channels are therefore used as wavelength-pair resources for distributing polarisation entanglement. Each DEMUX output is first connected to a fibre polarisation controller (FPC) and then to a $192\times192$ optical fibre switch (OFS), enabling programmable routing of individual wavelength channels.

Different output multiplexing approaches were used to connect the q-ROADM to the six users. Passive multiplexers (MUXs) have lower insertion loss than programmable wavelength-selective switches (WSSs), but the supported wavelengths are fixed by the MUX passbands and require dedicated OFS ports. In this experiment, Alice and Bob were each connected through a $1\times16$ MUX, providing access to different selected sets of DWDM channels while avoiding excessive use of OFS ports. This design represents a trade-off between low-loss operation and wavelength-assignment flexibility.

Although WSSs have higher insertion loss, approximately 5~dB compared with 1.3--2.7~dB for the MUXs, they provide wavelength-agnostic and programmable multiplexing. This reduces the number of OFS ports required for flexible channel assignment. Two $4\times16$ WSSs (Finisar WaveShaper 16000S) were used: one associated with Chloe and David, and the other with Faye and Grant. For each WSS, 16 ingress ports were connected to the OFS, while two egress ports were used as client ports connected to the users. The WSSs were configured to multiplex the selected wavelength channels from the OFS to the designated users.

\begin{figure*}[h]
\centering
\includegraphics[width=\linewidth]{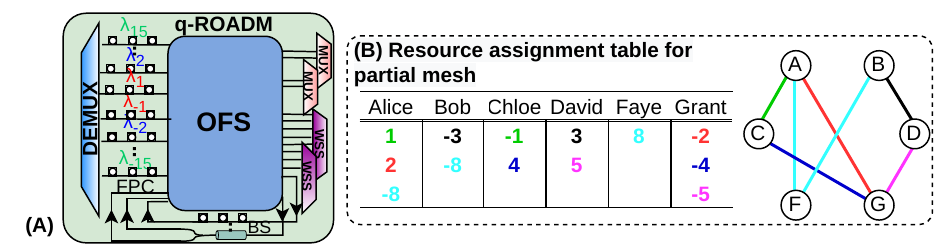}
\caption{q-ROADM architecture and partial-mesh wavelength assignment. (a) q-ROADM setup including DEMUX, FPCs, OFS, MUXs, WSSs and beam splitters. (b) Example partial-mesh resource assignment where wavelength channel $\lambda_{-8}$ is distributed through a 3~dB beam splitter.}
\label{fig:Q_ROADM_supplement}
\end{figure*}

In addition to the DEMUX, MUXs and WSSs, extra FPCs and beam splitters (BSs) were connected directly to the OFS to provide additional flexibility. By routing a wavelength channel $\lambda_i$ from the OFS to a BS, photons in that channel are probabilistically distributed to two output ports, and the channel can therefore be shared between two users. This allows a single-wavelength channel to serve multiple user links, increasing the network's effective connectivity.

An example partial-mesh wavelength-assignment scheme using a beam splitter is shown in \Cref{fig:Q_ROADM_supplement}b. In this configuration, wavelength channel $\lambda_{-8}$ is routed from the DEMUX output to the BS input. The two BS outputs are then connected to Alice and Bob through separate paths:
\begin{equation}
\mathrm{DEMUX}_{-8} \rightarrow \mathrm{BS}_{\mathrm{in}} \rightarrow \mathrm{BS}_{\mathrm{out},1} \rightarrow \mathrm{MUX}_{\mathrm{Alice}}
\end{equation}
and
\begin{equation}
\mathrm{DEMUX}_{-8} \rightarrow \mathrm{BS}_{\mathrm{in}} \rightarrow \mathrm{BS}_{\mathrm{out},2} \rightarrow \mathrm{FPC} \rightarrow \mathrm{MUX}_{\mathrm{Bob}}
\end{equation}
The FPC after the DEMUX output neutralises the fibre path to Alice, while the additional FPC in the second BS output path neutralises the fibre path to Bob. The subscript 0.5 used in the wavelength-assignment table denotes the 50:50 splitting ratio of the 3~dB BS.

This beam-splitter-assisted configuration can increase the number of supported user connections and provides an additional mechanism for scaling the network connectivity. In the present experiment, the number of users was limited to six by the available user modules and SNSPD channels, so configurations beyond six users were not investigated. Overall, the q-ROADM architecture provides a flexible and scalable platform for on-demand entanglement distribution, where wavelength channels can be routed, multiplexed, split and polarisation-neutralised according to the required network topology.

\subsection{SIAT Protocol Time Analysis}
\label{sec:siat_time_analysis}

\begin{figure*}[h]
\centering
\includegraphics[width=\linewidth]{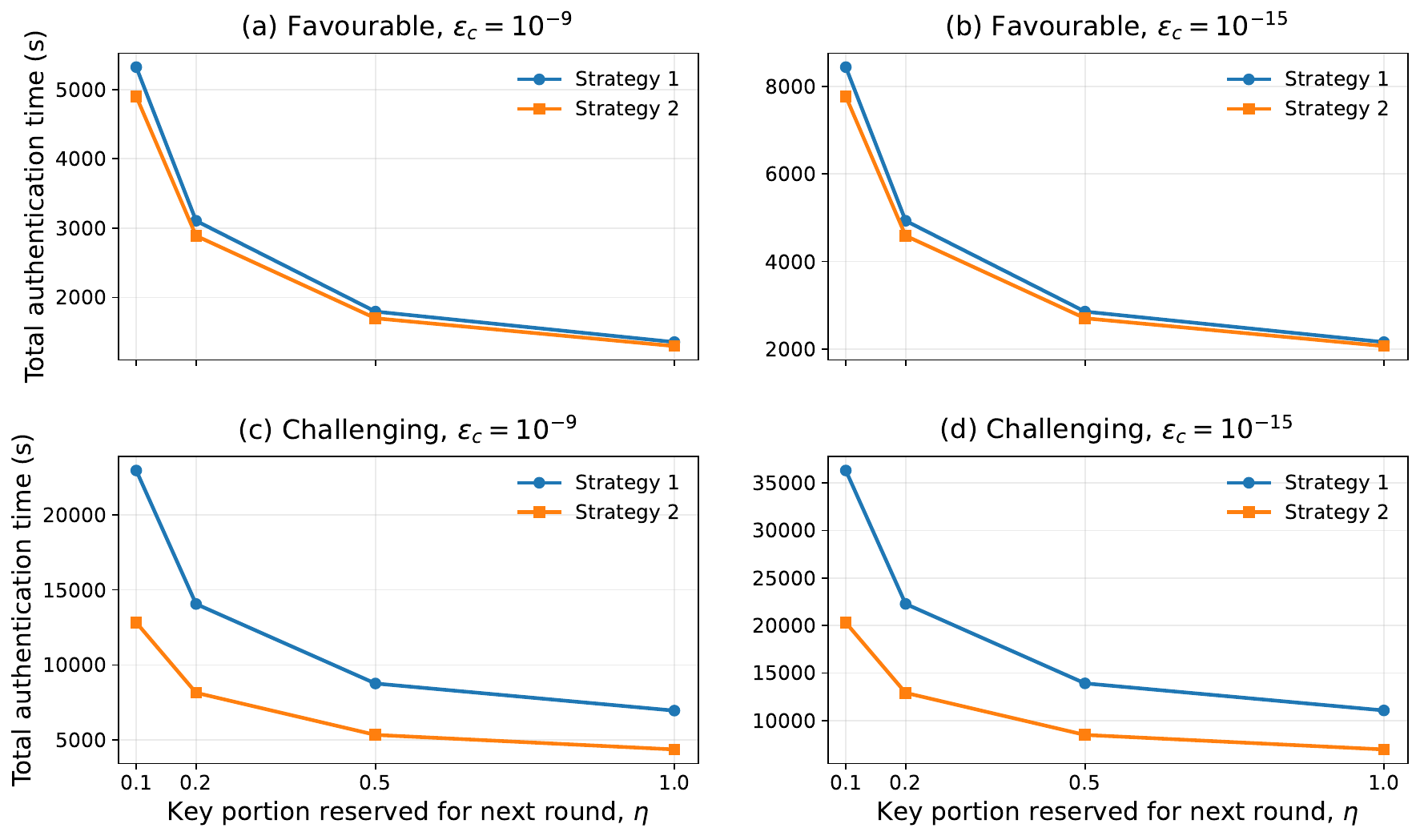}
\caption{Total SIAT authentication time for adding Alice to the five-user network as a function of the fraction of generated keys reserved for the next authentication round, $\eta$. Each data point is obtained by summing the Step~1 and Step~2 durations over all newly authenticated Alice--user links in the corresponding authentication strategy. Results are shown for S1 and S2 under favourable and challenging operating conditions, and for authentication insecurity values $\varepsilon_c$ of $10^{-9}$ and $10^{-15}$. The calculation uses link-dependent classical-to-secret-key ratios derived from the measured post-processing data. Lower $\eta$ leaves more generated key material available for other applications, but increases the direct Step~2 key-generation time.}
\label{fig:supp_siat_strategy_eta}
\end{figure*}

This section provides the time analysis for the SIAT protocol considered in the main text. We consider the scenario where a new user, Alice, is added to an existing five-user fully connected entanglement-based quantum network containing Bob, Chloe, David, Faye and Grant. The objective is to authenticate Alice with all existing users so that direct quantum communication links can subsequently be established without relying on the initially trusted intermediary users. Two authentication strategies are analysed. In Strategy~1, denoted as S1, Alice initially shares authentication keys with Bob and is subsequently authenticated with Chloe, David, Faye and Grant. In Strategy~2, denoted as S2, Alice initially shares authentication keys with Grant and is subsequently authenticated with Faye, David, Chloe and Bob.

The SIAT protocol can be described using three phases. In Phase~1, a trusted agency or trusted intermediate user distributes authentication keys to two parties that wish to establish a quantum communication link. In the present network, this initial trusted user is Bob for S1 and Grant for S2. In Phase~2, the transferred authentication keys are used to authenticate the first direct quantum communication session between Alice and the newly authenticated existing user. Here, ``newly authenticated'' refers to Alice's perspective: the user is already part of the five-user network, but has not yet established an authenticated direct link with Alice. In Phase~3, trust in the original intermediary is no longer required. Instead, Alice and the newly authenticated user use a portion, or all, of the keys generated in the previous direct quantum communication round to authenticate future rounds. Phase~3 is then repeated periodically, or according to application-specific security criteria, until the authenticated connection is no longer required.

For compact notation, let $P_1$, $P_2$ and $P_3^{(m)}$ denote Phase~1, Phase~2 and the $m$-th repetition of Phase~3, respectively. The SIAT process for each newly authenticated Alice--$X$ pair is written as
\begin{equation}
P_1 \rightarrow P_2 \rightarrow P_3^{(1)}
\rightarrow P_3^{(2)} \rightarrow \cdots \rightarrow P_3^{(m)} 
\label{eq:supp_siat_phases}
\end{equation}
The time between phases depends on the authentication security requirement, the fraction of generated key material reserved for future authentication, and the key-generation rate of the corresponding entanglement links.

For the time analysis, the transitions between phases are decomposed into two key-accumulation steps. Step~1 corresponds to the time required to accumulate sufficient authentication key material through the trusted intermediary or through flooding-based authentication-transfer paths, enabling the transition from Phase~1 to Phase~2. Step~2 corresponds to the time required for Alice and the newly authenticated user to generate sufficient fresh secret keys over their direct entanglement link, enabling the transition from Phase~2 to Phase~3. These two times are denoted as $T_{\mathrm{stp1}}$ and $T_{\mathrm{stp2}}$, respectively:
\begin{equation}
T_{P_1\rightarrow P_2}=T_{\mathrm{stp1}},
\qquad
T_{P_2\rightarrow P_3^{(1)}}=T_{\mathrm{stp2}} 
\label{eq:supp_siat_phase_steps}
\end{equation}
Each further Phase~3 refresh requires another direct key-generation period,
\begin{equation}
T_{P_3^{(m)}\rightarrow P_3^{(m+1)}} =
T_{\mathrm{stp2}}^{(m+1)},
\qquad m=1,2,\ldots 
\label{eq:supp_siat_phase3_refresh}
\end{equation}

We follow the authentication-key estimation used in Refs.~\cite{wegman1981new,solomons2022scalable}, where the number of secret bits required for Wegman--Carter authentication depends on the amount of authenticated classical communication exchanged during a QKD round. Let $a$ be the final secret-key length generated in one QKD round, and let $\varepsilon_c$ be the desired authentication insecurity. The amount of authenticated classical communication is written as
\begin{equation}
d = g a ,
\label{eq:supp_classical_bits}
\end{equation}
where $g$ is the classical-to-secret-key ratio, i.e., the number of authenticated classical bits exchanged per final secret-key bit. In Ref.~\cite{solomons2022scalable}, this ratio was estimated from the number of photon detection events and the classical information required for time tagging, basis announcement, error correction and status checks. In our analysis, the ratio is obtained from the measured post-processing data of each Alice--$X$ connection. For a link $\ell$, the ratio is defined as
\begin{equation}
g_{\ell}
=
\frac{D_{\ell}}{K_{\ell}}
=
\frac{D_{\ell}}{R_{\ell}\tau}
\label{eq:supp_link_ratio}
\end{equation}
Here, $D_{\ell}$ is the total number of authenticated classical bits exchanged during the measurement for link $\ell$, $K_{\ell}$ is the number of generated secret bits, $R_{\ell}$ is the measured SKR of link $\ell$, and $\tau$ is the integration time.

As an example, consider S1 when Alice is authenticated with Chloe. The measured single-count rates are 60,771~counts/s for Alice and 84,235~counts/s for Chloe. At this stage, Alice receives two channels, corresponding to the initial link with Bob and the new link with Chloe, while Chloe receives five channels. Considering only the time-tagging contribution, and assuming each time tag is encoded using a 64-bit number, the classical communication rate for exchanging time-tagging information is
$
D_{A-C}^{\mathrm{TT}}
=
(60771 + 84235)\times 64
=
9280384~\mathrm{bits/s}
\label{eq:supp_ac_classical_rate}
$
The direct Alice--Chloe link has a measured SKR of 77.32~bits/s. The corresponding time-tagging contribution to the classical-to-secret-key ratio is therefore
$
g_{A-C}^{\mathrm{TT}}
=
\frac{D_{A-C}^{\mathrm{TT}}}{R_{A-C}}
=
\frac{9280384}{77.32}
=
120025.45 
\label{eq:supp_ac_ratio}
$
In the SIAT timing calculation, the ratio $g_{A-X}$ denotes the full classical-to-secret-key ratio used for the Alice--$X$ link, including time-tagging information and other authenticated classical communication required by the post-processing procedure.
When error correction and other authenticated classical communication are included, the same procedure gives the full link-dependent ratio $g_{A-C}$. However, the extra information needed for post-processing is not comparable to the information needed for time tagging exchange. Therefore, in this paper, we assume $g_{A-X} \approx g_{A-X}^{\mathrm{TT}}$ for all the quantum links. This ratio $g_{A-X}$  is link-dependent, capturing differences in link loss, QBER, detector count rate, and secret key generation, and therefore provides a more accurate estimate of the authentication cost than using a fixed overestimated value for all user pairs.

The authentication requirement is determined by the amount of key that must be reserved for authenticating the next quantum communication round. We denote this reserved authentication key length by $a_{\mathrm{auth}}$. If a fraction $\eta$ of the final secret key generated in the current round is reserved for future authentication, then the total final key length that must be generated in the current round is $a_{\mathrm{auth}}/\eta$. Therefore, the corresponding amount of authenticated classical communication is:
\begin{equation}
d
=
g_{A-X}\frac{a_{\mathrm{auth}}}{\eta}
=
\frac{g_{A-X}}{\eta}a_{\mathrm{auth}}
\label{eq:supp_eta_classical_bits}
\end{equation}
where $g_{A-X}$ is the classical-to-secret-key ratio for the Alice--$X$ link when the full generated key length is considered.

The authentication key length required for the next round is then calculated as:
\begin{equation}
\begin{aligned}
L_{\mathrm{auth}}&(a_{\mathrm{auth}},\varepsilon_c,g_{A-X},\eta) \\
&=
4
\left[
\log_2\left(\frac{2}{\varepsilon_c}\right)
+
\log_2\log_2\left(\frac{g_{A-X}a_{\mathrm{auth}}}{\eta}\right)
\right] \\
&\quad \times
\log_2\left(\frac{g_{A-X}a_{\mathrm{auth}}}{\eta}\right).
\end{aligned}
\label{eq:supp_auth_key_length_eta}
\end{equation}
The reserved authentication key must be at least equal to the authentication key required for the next round. Therefore, $a_{\mathrm{auth}}$ is obtained from the self-consistency condition:
\begin{equation}
a_{\mathrm{auth}}
=
L_{\mathrm{auth}}(a_{\mathrm{auth}},\varepsilon_c,g_{A-X},\eta)
\label{eq:supp_eta_condition}
\end{equation}
Given the insecurity parameter $\varepsilon_c$, the link-dependent classical-to-key ratio $g_{A-X}$, and combining Eq.\ref{eq:supp_auth_key_length_eta} and Eq.\ref{eq:supp_eta_condition}, we can calculate the required total key length and authentication key lengths required for implementing SIAT protocol.

In this work, we evaluate two representative authentication insecurity values, $\varepsilon_c=10^{-9}$ and $\varepsilon_c=10^{-15}$, and different fractions, $\eta$, of keys generated per round used as the authentication keys in the next round. A smaller value of $\varepsilon_c$ corresponds to a stricter authentication requirement. A smaller value of $\eta$ means that a smaller fraction of the generated key is reserved for future authentication, leaving more key material available for other quantum applications or protocols.

The same Phase~3 authentication requirement determines the key material that must be accumulated in both Step~1 and Step~2 for a given Alice--$X$ authentication process. Step~1 accumulates the reserved authentication key through the trusted intermediary or flooding-based authentication-transfer paths. Therefore, the Step~1 duration is
\begin{equation}
T_{\mathrm{stp1}}^{A-X}
=
\frac{a_{\mathrm{auth}}}
{R_{\mathrm{stp1}}^{A-X}}
\label{eq:supp_siat_step1_time}
\end{equation}
where $R_{\mathrm{stp1}}^{A-X}$ is the effective secret key rate available through the Step~1 authentication-transfer process. For flooding-based SIAT in the security-enhancing mode, authentication keys transferred through multiple non-overlapping paths are XORed. In this case, each selected path must accumulate enough key material before the XORed authentication key can be formed. Therefore, the effective Step~1 rate is limited by the slowest selected path:
\begin{equation}
R_{\mathrm{stp1}}^{A-X}
=
\min_{p\in\mathcal{P}_{A-X}} R_p
\label{eq:supp_siat_xor_rate}
\end{equation}
Here, $\mathcal{P}_{A-X}$ is the set of selected flooding paths used to authenticate the Alice--$X$ connection, and $R_p$ is the end-to-end secret key rate of path $p$. For a multi-hop path, the end-to-end rate is limited by the weakest link along that path:
\begin{equation}
R_p
=
\min_{\ell\in p} R_{\ell}
\label{eq:supp_path_rate}
\end{equation}

After the Step~1 authentication transfer, Step~2 establishes the direct Alice--$X$ entanglement link and generates fresh keys for future authentication. Since only a fraction $\eta$ of the generated key is reserved for the next authentication round, the total final key length required in Step~2 is $a_{\mathrm{auth}}/\eta$. The Step~2 duration is therefore
\begin{equation}
T_{\mathrm{stp2}}^{A-X}
=
\frac{a_{\mathrm{auth}}/\eta}{R_{A-X}}
=
\frac{a_{\mathrm{auth}}}{\eta R_{A-X}}
\label{eq:supp_siat_step2_time}
\end{equation}
where $R_{A-X}$ is the direct SKR of the Alice--$X$ link. In the main-text analysis, we take $\eta=1$, meaning that all generated keys in the direct round are reserved for authenticating the next round. Smaller values of $\eta$ correspond to reserving only a portion of the generated keys for future authentication, leaving the remaining key material available for other quantum applications or protocols.

\begin{figure*}[!h]
\centering
\includegraphics[width=\linewidth]{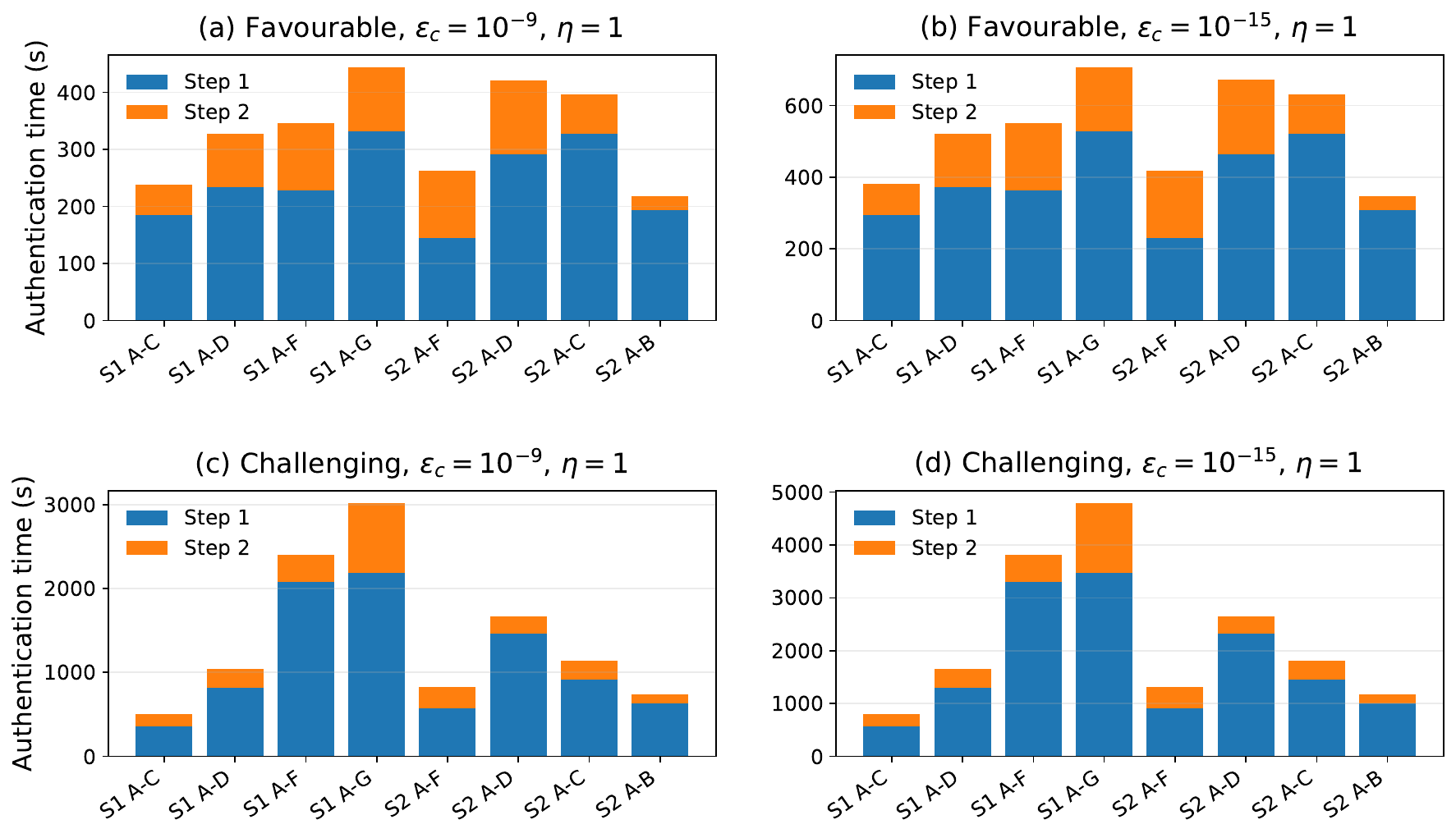}
\caption{Link-level SIAT timing for $\eta=1$, where all generated keys are reserved for the next authentication round. Each bar shows the time required for one newly authenticated Alice--$X$ link, decomposed into Step~1 and Step~2. Results are shown for favourable and challenging operating conditions and for authentication insecurity values $\varepsilon_c=10^{-9}$ and $\varepsilon_c=10^{-15}$.}
\label{fig:supp_siat_link_timing}
\end{figure*}

The total time required to authenticate Alice with all existing users is obtained by summing the Step~1 and Step~2 durations for each newly authenticated Alice--$X$ pair. For strategy 1 (S1), where Alice is authenticated in the order $C,D,F,G$, the total authentication time is
\begin{equation}
T_{\mathrm{S1}}
=
\sum_{X\in\{C,D,F,G\}}
\left(
T_{\mathrm{stp1}}^{A-X}
+
T_{\mathrm{stp2}}^{A-X}
\right)
\label{eq:supp_siat_strategy1_time}
\end{equation}
For strategy 2 (S2), where Alice is authenticated in the order $F,D,C,B$, the total authentication time is
\begin{equation}
T_{\mathrm{S2}}
=
\sum_{X\in\{F,D,C,B\}}
\left(
T_{\mathrm{stp1}}^{A-X}
+
T_{\mathrm{stp2}}^{A-X}
\right)
\label{eq:supp_siat_strategy2_time}
\end{equation}
The two strategies use the same physical q-ROADM network, but differ in the order in which Alice establishes authenticated links with the existing users. Since each Step~1 process may use different trusted-node or flooding paths, and each Step~2 process depends on a different SKR of the direct Alice--$X$ link, the total authentication time depends on the authentication order and on the quantum network performance, including system loss, detector timing jitter, source heralding efficiency and fibre length.

After Alice has been authenticated with a given user $X$, the Phase~3 refresh can be repeated for that Alice--$X$ connection. If the direct link condition remains approximately stable, the time required to initialise SIAT and complete $m$ refresh rounds for this pair is
\begin{equation}
T_{A-X}^{(m)}
\approx
T_{\mathrm{stp1}}^{A-X}
+
m T_{\mathrm{stp2}}^{A-X}
\label{eq:supp_siat_pair_refresh_time}
\end{equation}
More generally, if the link rate varies between refresh rounds, the total time is
\begin{equation}
T_{A-X}^{(m)}
=
T_{\mathrm{stp1}}^{A-X}
+
\sum_{k=1}^{m}
T_{\mathrm{stp2}}^{A-X,k}
\label{eq:supp_siat_pair_refresh_time_general}
\end{equation}

The time analysis reported in the main text is calculated using the measured asymptotic SKRs and the link-dependent classical-to-secret-key ratios of the dynamically configured entanglement links. The favourable case uses approximately 15\% source heralding efficiency and low detector jitter below 100~ps. The challenging case uses approximately 3\% source heralding efficiency and detector jitter of 300--350~ps. During Step~1, the q-ROADM establishes the trusted-node or flooding paths required for authentication transfer. During Step~2, the network is reconfigured to establish the direct entanglement link between Alice and the newly authenticated user. In the main-text timing comparison, we take $m=1$, meaning that at least one direct Step~2 round is completed so that future authentication no longer requires multi-path trusted-node transfer.

This dynamic operation is essential for efficient protocol execution. Under favourable conditions, where most links have relatively high SKR, the two authentication strategies give similar total authentication times. Under challenging conditions, however, low heralding efficiency, higher timing jitter and higher accidental-coincidence rates reduce the SKR of some links significantly. The slowest flooding path can then dominate $T_{\mathrm{stp1}}$, and the direct Alice--$X$ SKR can dominate $T_{\mathrm{stp2}}$. As a result, the authentication order becomes important, and one strategy can complete significantly faster than another. This explains the strategy-dependent SIAT timing reported in the main text and demonstrates how the q-ROADM can support protocol-aware entanglement distribution by adapting the network topology to the required SIAT step.

As shown in \Cref{fig:supp_siat_strategy_eta}, the total SIAT authentication time increases for stricter authentication security and for smaller key-reservation fractions. The use of link-dependent ratios leads to different timing penalties for different Alice--$X$ links, because each link has a different combination of SKR and classical-communication overhead. The link-level timing in \Cref{fig:supp_siat_link_timing} shows that the quickest total authentication time ($\eta=1$) can be dominated by a small number of weak quantum links, especially under challenging network conditions, while the higher security requirement will require a longer waiting time. This supports the use of the q-ROADM as a protocol-aware entanglement distribution layer, where the network topology and authentication order can be adapted according to the actual link conditions.

\end{document}